\documentclass[]{spie}  

\usepackage{amsmath,amsfonts,amssymb}
\usepackage{graphicx}
\usepackage[colorlinks=true, allcolors=blue]{hyperref}

\title{High-speed, low-noise, multi-megapixel CCDs for next generation X-ray observatories}

\author[a,b]{Haley R. Stueber}
\author[a]{Tanmoy Chattopadhyay}
\author[a]{Peter Orel}
\author[a,b,c]{Steven W. Allen}
\author[d]{Marshall W. Bautz}
\author[e]{Michael Cooper}
\author[e]{Kevan Donlon}
\author[d]{Catherine E. Grant}
\author[a]{Sven Herrmann}
\author[d]{Jill Juneau}
\author[d]{Beverly J. LaMarr}
\author[e]{Christopher Leitz}
\author[d]{Eric D. Miller}
\author[a,c]{R. Glenn Morris}
\author[a]{Declan O'Neill}
\author[a,b]{Abigail Y. Pan}
\author[a]{Tonya L. Peshel}
\author[a]{Artem Poliszczuk}
\author[d]{Gregory Y. Prigozhin}
\author[e]{Keith Warner}
\affil[a]{Kavli Institute for Particle Astrophysics and Cosmology, Stanford University, 452 Lomita Mall, Stanford, CA 94305, USA}
\affil[b]{Department of Physics, Stanford University, 382 Via Pueblo Mall, Stanford CA 94305, USA}
\affil[c]{SLAC National Accelerator Laboratory, 2575 Sand Hill Road, Menlo Park, CA 94025, USA}
\affil[d]{MIT Kavli Institute for Astrophysics and Space Research, Massachusetts Institute of Technology, 70 Vassar St, Cambridge, MA 02139, USA}
\affil[e]{MIT Lincoln Laboratory, 244 Wood St building 1324, Lexington, MA 02421, USA}

\authorinfo{Further author information: (Send correspondence to H.R.S.)\\H.R.S.: E-mail: hstueber@stanford.edu, Telephone: 1 920 252 5189}

\begin{document} 
\maketitle

\begin{abstract}
Next generation X-ray observatories require fast, low-noise, low-power, multi-megapixel imaging spectrometers. To meet these demands, the X-ray Astronomy and Observational Cosmology (XOC) Group at Stanford, in partnership with the MIT Kavli Institute and MIT Lincoln Laboratory (MIT-LL), is developing multi-channel X-ray charge-coupled devices (CCDs) and fast readout architectures. We report the energy resolution and noise performance achieved with a full-scale (1440x1440-pixel), 16-channel, front-illuminated MIT-LL CCD detector developed for the Advanced X-ray Imaging Satellite (AXIS) concept, the CCID-100, read out using two Multi-Channel Readout Chip (MCRC) V1 application-specific integrated circuit (ASIC) chips in the new Stanford CCID-100 test setup. We describe an automated method for bias optimization on each CCD channel, and integrated debugging features of the front-end ASIC and readout system. The demonstrated performance confirms that these systems can meet the speed and noise requirements of future strategic X-ray missions. 
\end{abstract}

\keywords{X-ray, CCDs, Low Noise, Fast Readout, AXIS}

\section{INTRODUCTION}
\label{sec:intro} 

\noindent Astrophysical X-rays trace some of the hottest, most extreme environments in the universe. The next generation of strategic, wide-field X-ray imaging missions will serve as windows to peer into these extreme landscapes. These instruments will require megapixel detectors capable of fast readout speeds to avoid pileup when resolving bright, energetic X-ray sources such as active galactic nuclei (AGN), but also maintain low noise levels to suppress backgrounds when observing the diffuse, low-luminosity universe.

Many device candidates meet either the noise or the speed performance metrics of such a mission. Active pixel sensors (APS) such as Hybrid CMOS detectors (HCDs) \cite{HCMOS07, HCMOS17}, while capable of delivering fast readout speeds, are often limited by read noise\cite{chattopadhyay18_HCDoverview}. Monolithic CMOS detectors (MCMOS)\cite{Kenter2018SPIE10762E..09K} can deliver on both speed and noise performance, but lack sensitivity at energies larger than 6\,keV. Depleted Field Effect Transistor (DEPFET) devices, like those onboard NewAthena's wide-field imager\cite{athenaSPIE2017}, also deliver on speed and noise, but their typically larger pixel sizes would be less desirable for an observatory targeting high spatial resolution. 

X-ray charge-coupled devices (CCDs) deliver many of the critical components of a high-resolution, wide-field X-ray observatory -- small pixel sizes, low noise, and larger depletion depths -- but they don't yet achieve the necessary frame rates.
The X-ray Astronomy and Observational Cosmology (XOC) group at Stanford, in collaboration with the Massachusetts Institute of Technology Kavli Institute (MKI) and MIT Lincoln Laboratory (MIT-LL), have been developing and testing fast, low noise, large-format CCDs to address this technology gap and support next generation, strategic X-ray mission concepts. 

In this manuscript, we present initial results obtained with two MIT-LL CCID-100 devices tested at Stanford. Other aspects of the performance of two additional CCID-100s, such as the noise behavior and charge transfer characteristics, can be found in [\citenum{Prigozhin2026}] and [\citenum{LaMarr2026}]. These 16-channel, multi-megapixel X-ray CCD prototypes were originally developed for the Advanced X-ray Imaging Satellite (AXIS) mission concept\cite{chrisSPIE2023, millerAXIS2023}. Designed to achieve that mission's targeted sub-arcsecond resolution while maintaining excellent energy resolution, low $<3$\,electron noise, and fast, 5\,frames/s readout speeds, these detectors and their associated readout electronics form essentially mission-ready devices suitable for future X-ray missions ranging from Medium-Class Explorers (MIDEX) to flaglets and Probe-class missions, and provide a stepping stone towards a camera for a future X-ray flagship observatory.

This manuscript is organized as follows: In Section~\ref{sec:CCID100}, we describe the CCID-100 architecture and its low-noise, high-speed output stage design. Section~\ref{sec:test_setup} and subsections therein outline the Stanford beamline test system and readout electronics used to test and characterize the CCID-100 devices, as well as discuss our fast, parallelized, application-specific integrated circuit (ASIC)-based readout scheme and associated system monitoring and diagnostic tools. In Section~\ref{sec:bias_scan}, we present the noise characteristics of the CCID-100 devices tested, as well as a method for efficiently finding an optimal output stage bias point that minimizes the average read noise across detector channels. In Section~\ref{sec:results}, we present the single pixel event spectra obtained across all channels of two CCID-100 devices, which demonstrate the noise performance targeted by future strategic X-ray missions.

\section{MIT-LL CCID-100}
\label{sec:CCID100} 

\begin{figure} [ht!]
   \begin{center}
   \begin{tabular}{c}
   \includegraphics[height=7cm]{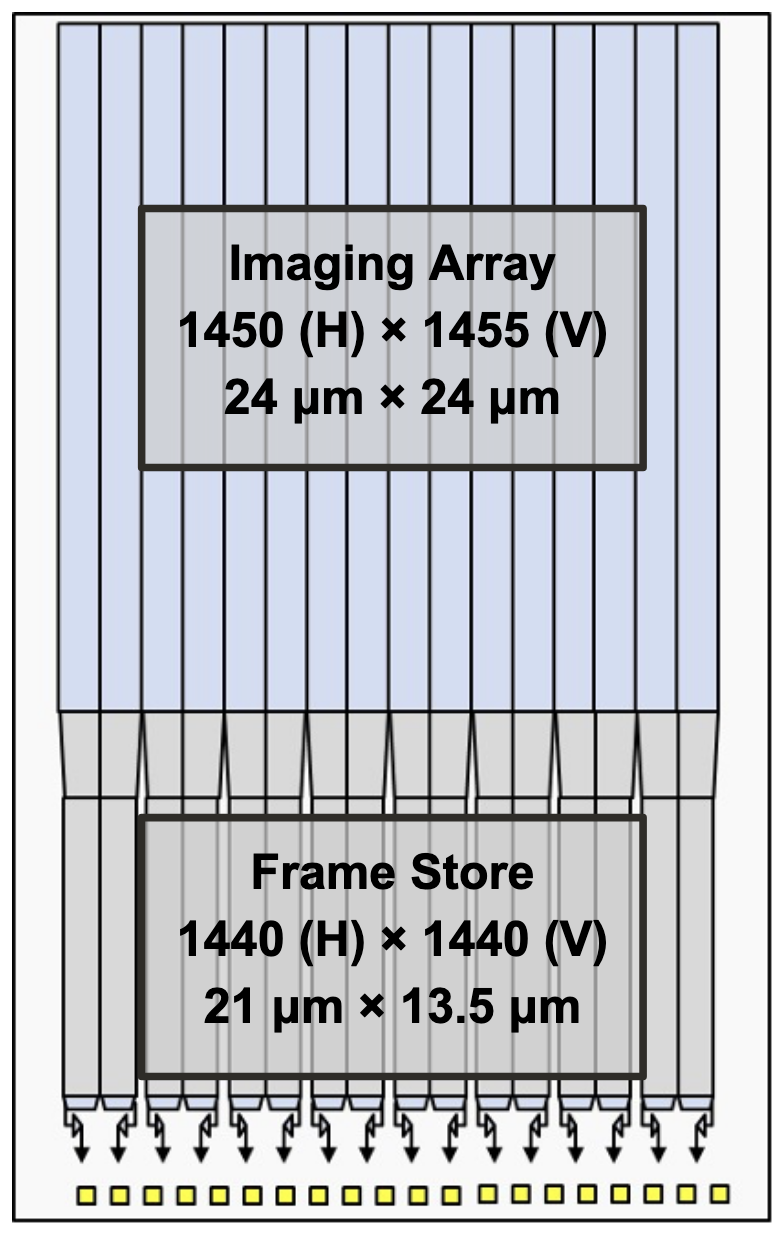}
   \includegraphics[height=7cm]{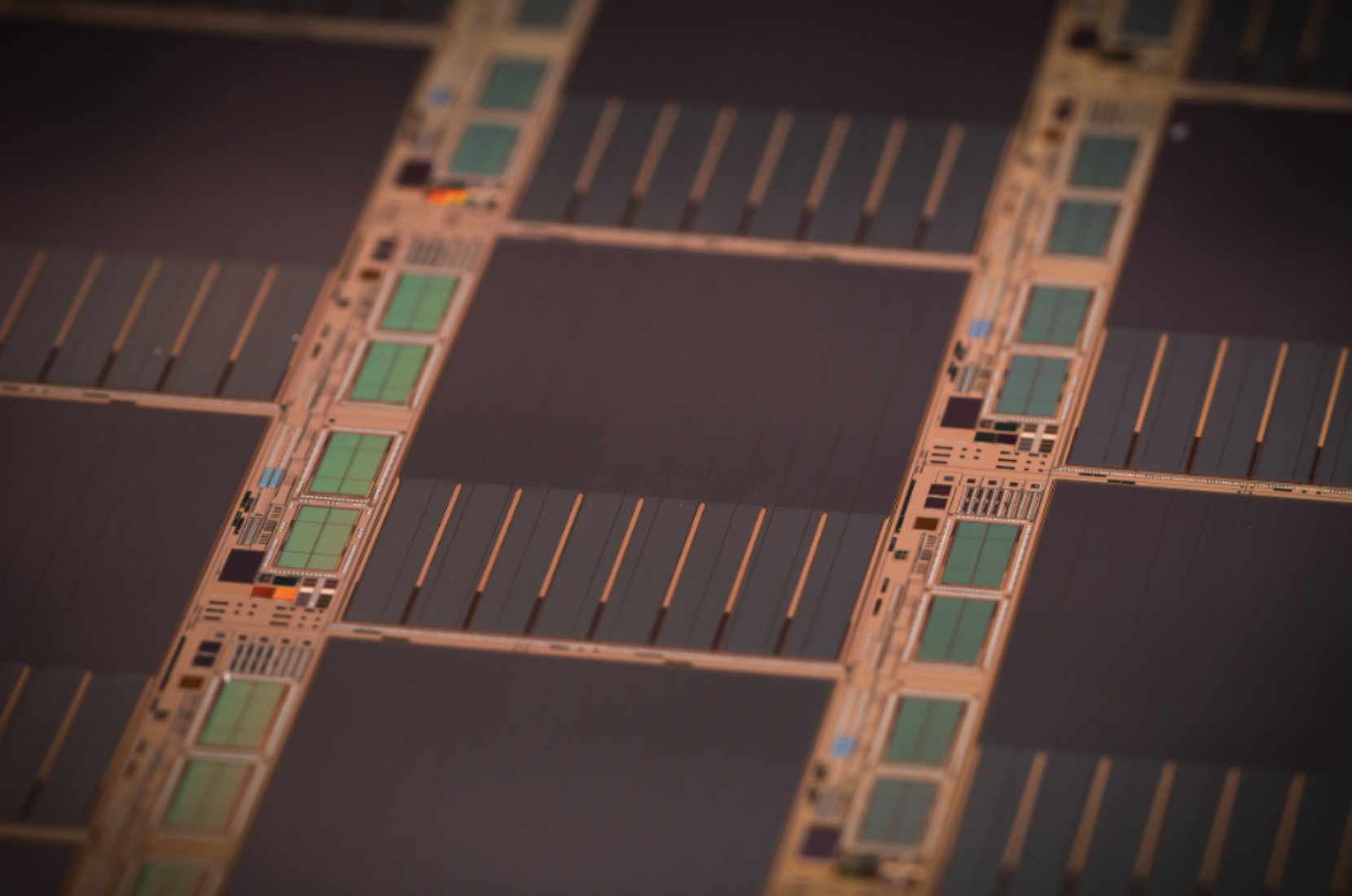}
   \end{tabular}
   \end{center}
   \caption[] 
   { \label{fig:CCID100} 
   {\it Left:} Schematic of the CCID-100 layout, depicting an imaging area of $1450\times1455$ $24\times24\,\rm \mu m$ pixels, and a frame store consisting of $1440\times1440$ $21\times13.5\,\rm \mu m$ pixels. {\it Right:} Fabricated, front-illuminated CCID-100 chips on a wafer.
}
\end{figure}

\noindent The MIT-LL CCID-100 is a $1440\times1440$, $24\,\rm \mu m$ pixel, 16-channel device (Fig.~\ref{fig:CCID100}). It benefits from single-level polysilicon process fabrication technology that enables low power operation of the CCD and fast transfer speeds. A fabricated CCID-100 wafer is pictured on the right of Figure~\ref{fig:CCID100}.

\begin{figure} [ht!]
   \begin{center}
   \begin{tabular}{c}
   \includegraphics[height=7cm]{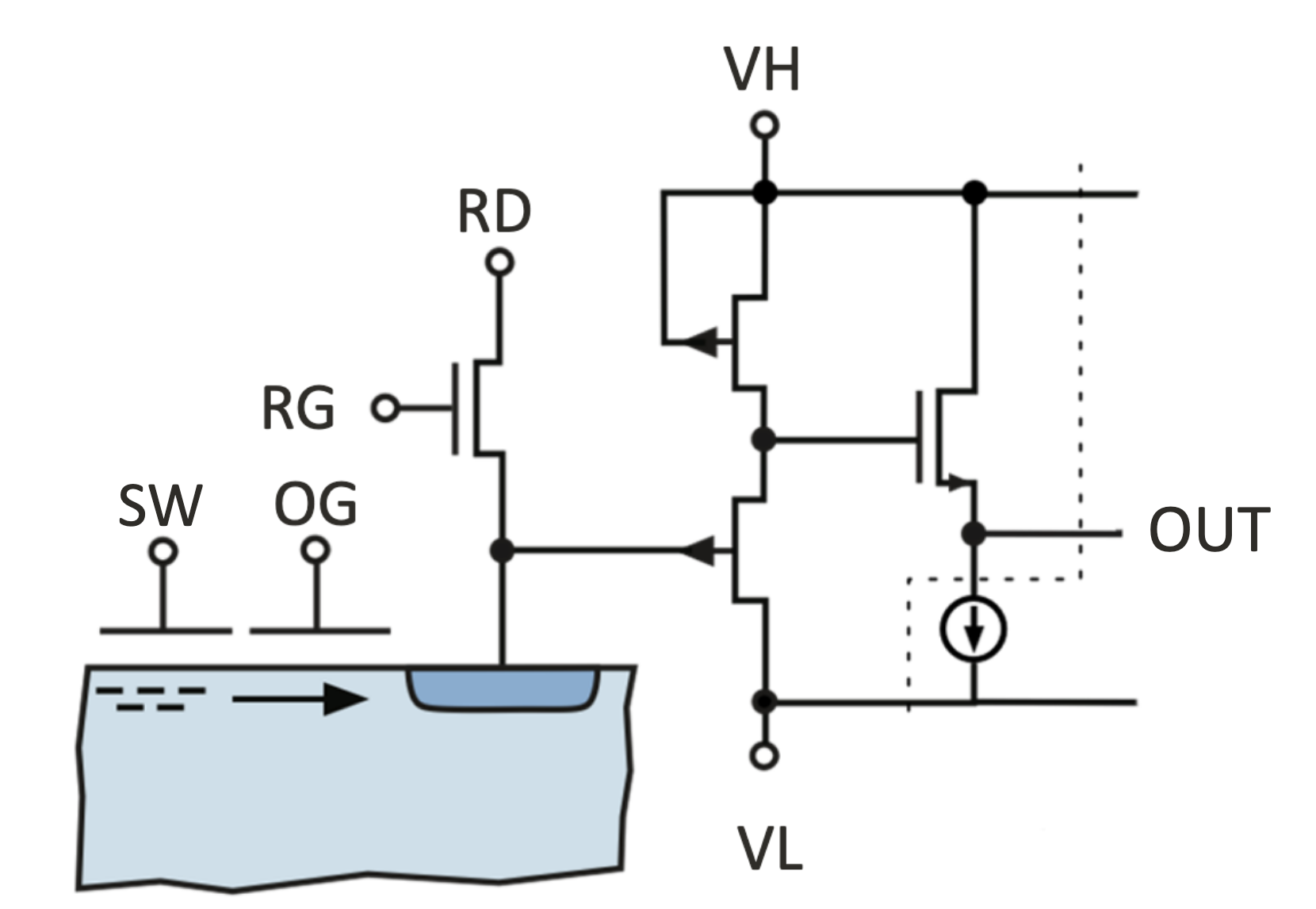}
   \end{tabular}
   \end{center}
   \caption[] 
   { \label{fig:output} 
   CCID-100 two-stage output schematic diagram. The first stage consists of a pJFET while the second stage consists of an nMOSFET. The summing well (SW), output gate (OG), reset gate (RG), reset drain (RD), VH (JFET current source bias) and VL (JFET drain bias) nodes are labeled.
}
\end{figure}

The CCID-100 features a two-stage output consisting of a fast, high-conversion gain source-follower p-channel junction field effect transistor (pJFET) followed by a large-bandwidth n-channel metal oxide semiconductor field effect transistor (nMOSFET). An output stage schematic diagram is presented in Figure~\ref{fig:output}\cite{tanmoyJATIS22}. The performance of this output stage architecture was previously demonstrated with the $512\times512$, $8\,\rm \mu m$ pixel CCID-93 devices, which achieved readout speeds of up to 5\,MPixel/s while maintaining read noise levels below 4 electrons \cite{Stueber10.1117}.

In this work, we present results obtained from two MIT-LL CCID-100, front-illuminated devices, referred to from this point on as Wafer 6 (W6) and Wafer 16 (W16). Although both are buried channel detectors that contain a subchannel implant for improved gain and speed performance, the W16 detector is a low-threshold variant, meaning that it has a lower energy source-drain channel implant, resulting in a channel that is located closer to the surface. This is expected to result in increased transconductance and lower noise levels\cite{Leitz2025}.

\section{TEST SETUP \& READOUT ELECTRONICS}
\label{sec:test_setup}

\noindent Testing, characterizing, and optimizing the 2-megapixel, 16-channel CCID-100 devices to achieve baseline performance metrics for missions like AXIS requires test systems and readout electronics conducive to fast, low-noise operation. In the following subsections, we describe the Stanford X-ray beamline CCID-100 test bed, designed to accommodate low temperature operation of the detectors in conjunction with custom readout electronics that leverage the fast, low-power performance capabilities of ASIC chips. We summarize the Stanford-developed multi-channel readout chip (MCRC) V1 CCID-100 readout ASIC architecture, operation modes, and performance specifications (see [\citenum{herrmann20_mcrc}], [\citenum{porelMCRCspie2022}], and [\citenum{porelMCRCspie2024}] for more details and latest performance). Finally, we describe newly integrated debugging features intended to streamline integration and assist in troubleshooting the detectors, test setup, and readout electronics.

\subsection{STANFORD CCID-100 BEAMLINE TEST SYSTEM}
\label{sec:beamline}

\begin{figure} [ht!]
   \begin{center}
   \begin{tabular}{c}
   \includegraphics[height=7cm]{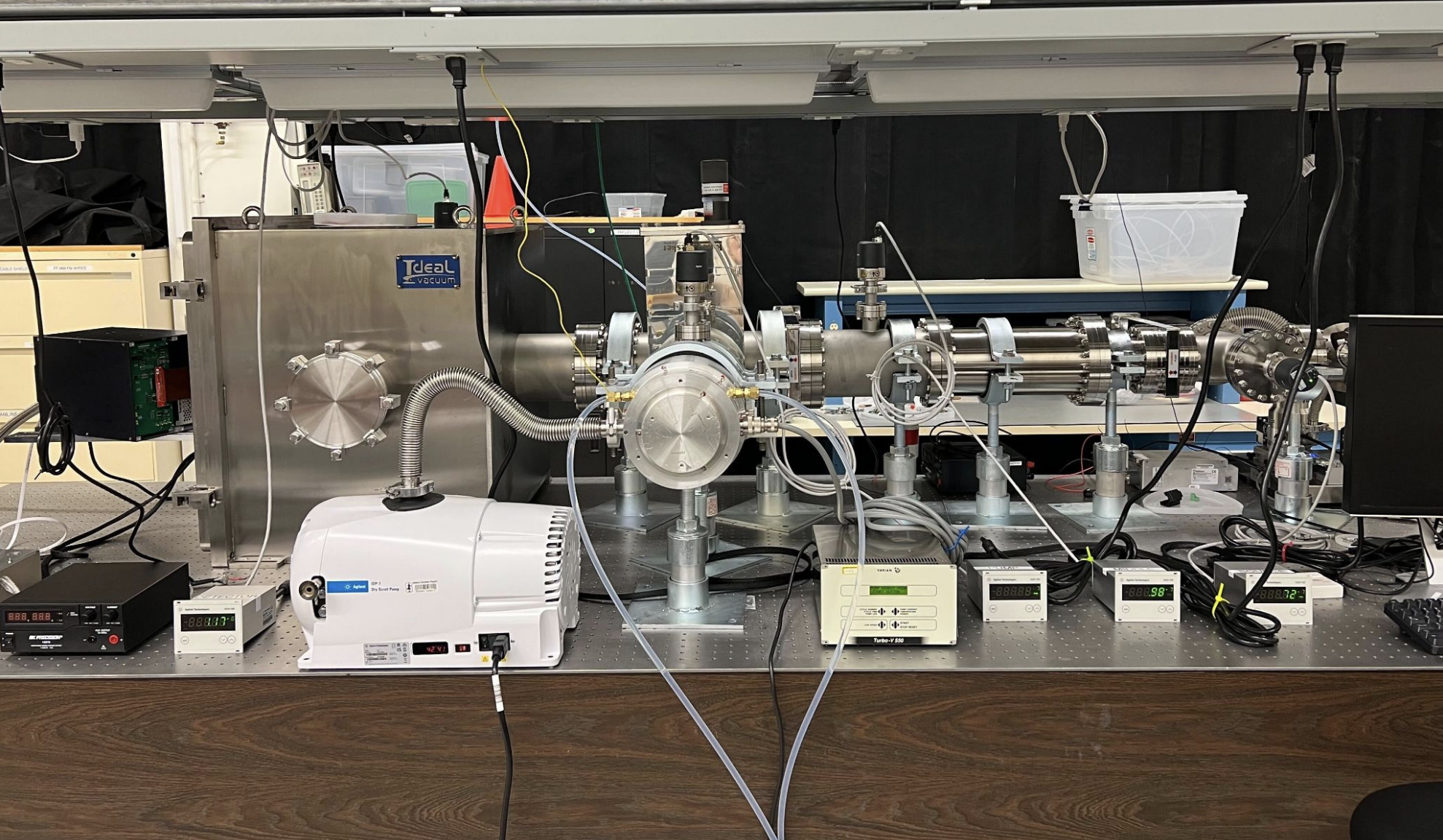}
   \end{tabular}
   \end{center}
   \caption[] 
   { \label{fig:test_setup} 
  XOC Gen 1.0 X-ray beamline. For details on the test setup, see [\citenum{10.1117/12.3017691}].
}
\end{figure}

\begin{figure} [ht!]
   \begin{center}
   \begin{tabular}{c}
   \includegraphics[height=5.5cm]{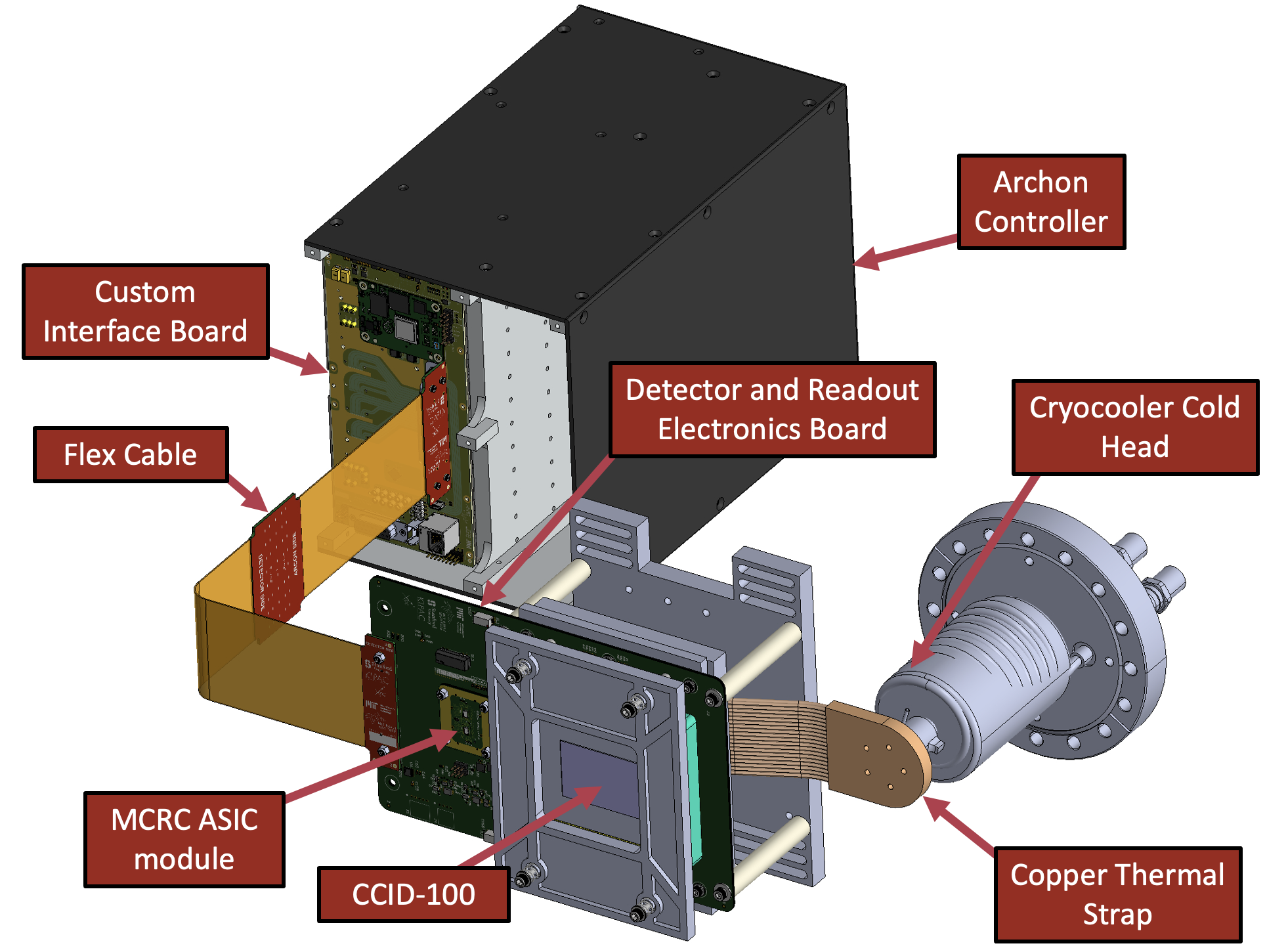}
   \includegraphics[height=5.5cm]{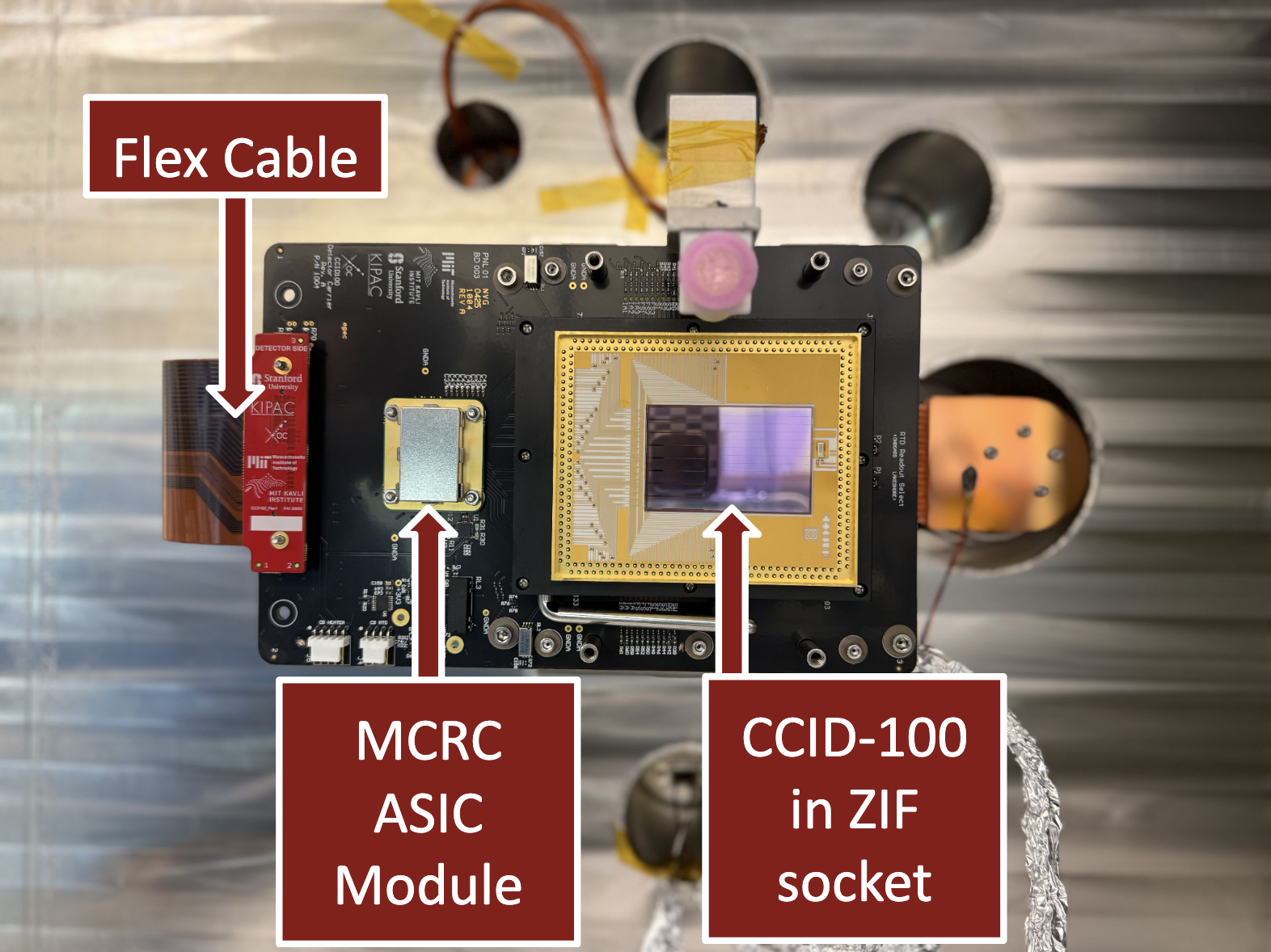}
   \end{tabular}
   \end{center}
   \caption[] 
   { \label{fig:mount} 
   {\it Left:} CAD model of the CCID-100 test assembly. Components include the Archon controller plus custom interface board, which contains a Raspberry Pi computer module, flex cable, MCRC ASIC and readout electronics, CCID-100 detector package and socket connector, cooling mask elements, and the cryocooler cold head and thermal strap. {\it Right:} CCID-100 with readout electronics, ASIC, and flex cable mounted in Gen 1.0 XOC X-ray Beamline.
}
\end{figure}

\noindent The Stanford Gen 1.0 XOC X-ray Beamline vacuum test system (Fig.~\ref{fig:test_setup}, see [\citenum{10.1117/12.3017691}] for full details on the beamline test system) was used to test and characterize the W6 and W16 detectors. As part of this system, a new printed circuit board (PCB) was designed to support the 16-channel CCID-100 detector coupled with ASIC readout. The right side of Figure~\ref{fig:mount} shows the new PCB mounted inside the beamline. The board integrates a large zero insertion force (ZIF) socket for detector insertion, ASIC readout, and a 200-pin connector. The latter connects a PCB flex cable that routes all of the analog and control signals from an Archon controller\footnote{http://www.sta-inc.net/archon/} situated outside of the beamline vacuum chamber to this PCB. This Archon controller is an all-in-one data acquisition (DAQ) pipeline back-end, which provides all of the necessary control signals for the detector and hosts the ADCs and the data processing pipeline. It has configurable hardware, and in our case, it is equipped with 16 ADCs running at 100 MSPs in parallel that sample the MCRC signal waveform. The large data stream is then fed into the onboard digital pulse processing (DPP) algorithm that finally extracts the image data, which is then presented to the user via a graphical user interface (GUI) on the host computer.

A new custom cooling mask setup was integrated for the CCID-100's, which includes a custom copper thermal strap from Technology Applications, Inc.\footnote{https://www.techapps.com/} for facilitating cooling via a cryocooler and ensuring good thermal contact to minimize heat loss at interfaces. With this setup, we can achieve stable detector temperatures of $-100^{\circ}\rm C$. The left side of Figure~\ref{fig:mount} shows a labeled CAD model of the full CCID-100 mounting and cooling setup, including the flex cable and Archon controller. The results presented in Section~\ref{sec:results} were obtained by illuminating the CCID-100 devices with an Fe-55 radioactive source mounted inside the beamline.

\subsection{STANFORD MCRC DUAL ASIC READOUT}
\label{sec:MCRC}

\begin{figure} [ht!]
   \begin{center}
   \begin{tabular}{c}
   \includegraphics[height=5.2cm]{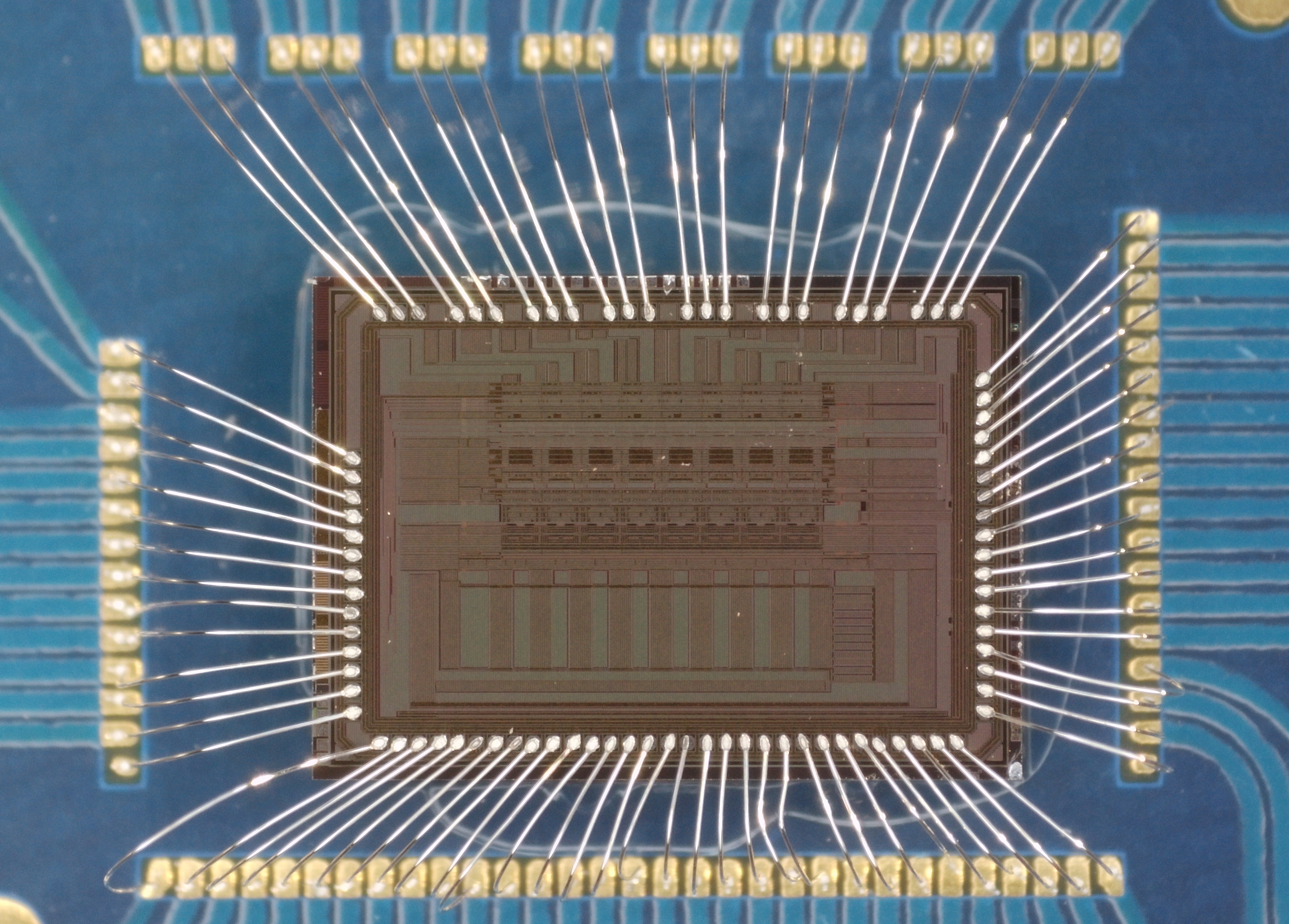}
   \includegraphics[height=5.2cm]{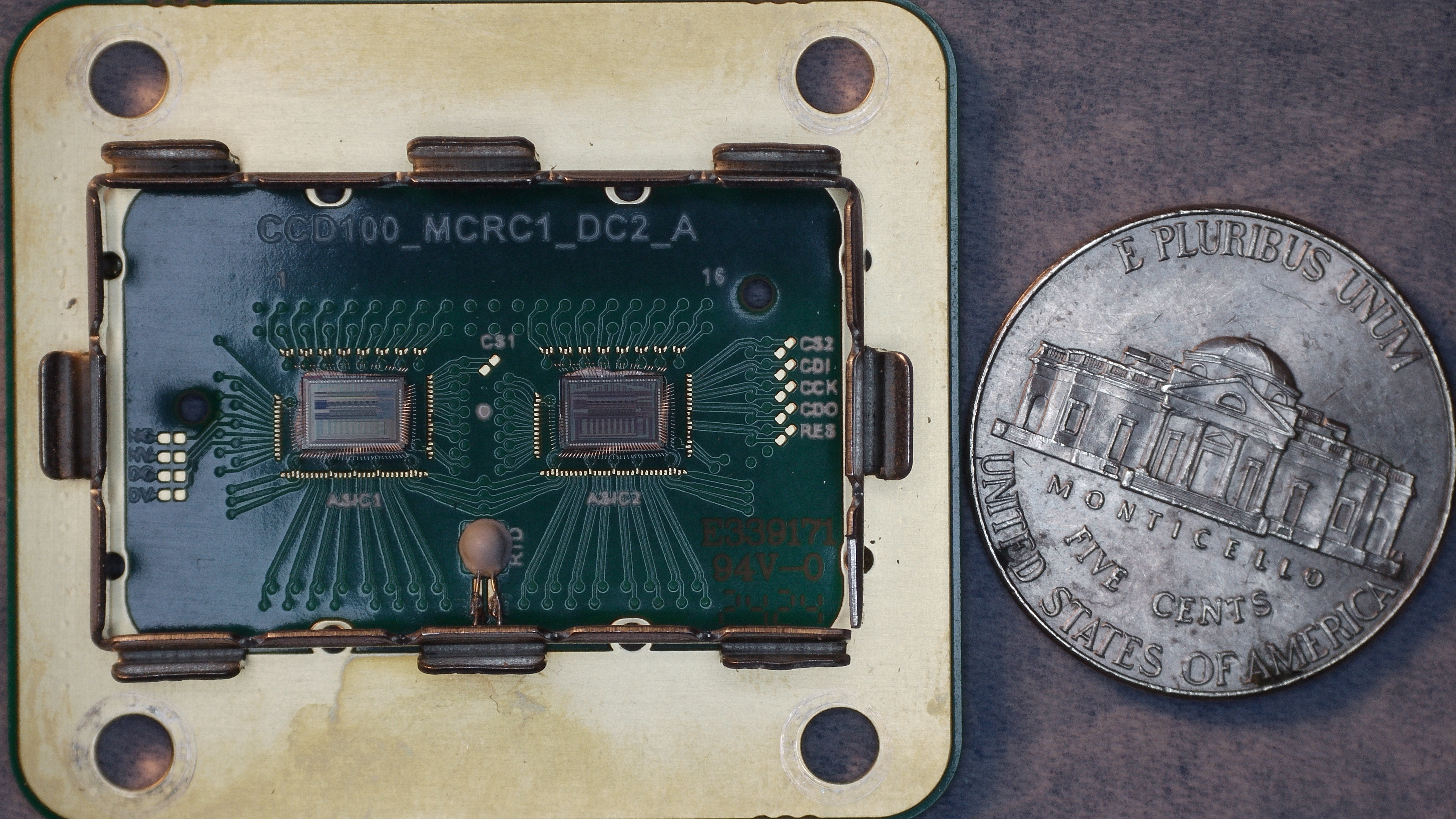}
   \end{tabular}
   \end{center}
   \caption[] 
   { \label{fig:ASIC} 
   {\it Left:} Microscope image of an MCRC V1 ASIC readout chip. Dimensions are $4160\,\rm \mu m \times 2900\,\rm \mu m$. {\it Right:} Dual MCRC V1 CCID-100 readout board, with a nickel for scale.
}
\end{figure}

\noindent The MCRC V1 has eight analog readout channels operating in parallel that can each be interfaced with a CCD output. The 16-channel CCID-100 requires two of these ASICs for full parallelization. A microscope image of the MCRC V1 is shown on the left of Figure~\ref{fig:ASIC} (dimensions of the chip $\rm 4160\,\mu m \times \rm 2900\,\mu m$), while the dual ASIC board is shown on the right with a nickel for scale.
Each of the ASIC channels features two user-selectable inputs that can be interfaced with the appropriate corresponding detector output: a high-impedance, source-follower input which also integrates a programmable current source to provide biasing to the CCD output, or a low-impedance drain readout input featuring an all-in-one integrated solution with a programmable current source, active cascode, and a current-to-voltage converter. The CCID-100 devices utilize the source-follower input, which is directly connected to a preamplifier that has user selectable gains of 8 or 16\,V/V. In addition, the preamplifier also converts the input single-ended signal to a differential format, providing better immunity to noise and cross-talk between adjacent channels. The preamplifier output is fed into a fully differential unity-gain output buffer, which can directly drive an analog-to-digital converter (ADC) without the need for external buffers, thus bringing down the system complexity and current draw. A digital serial peripheral interface (SPI) based on low voltage differential signaling (LVDS) protocol is used to access the ASIC and program its settings. This includes controlling internal switch logic and digital-to-analog converters (DACs) that supply biases to the internal analog circuitry. 

Physically and functionally compact, the MCRC V1 ASIC simplifies CCD board design by reducing the physical footprint of the readout electronics as well as the number of discrete components required. It has a large bandwidth, high input and output dynamic ranges, and outperforms the rate and noise capabilities of our best discrete readout solutions, contributing less than one electron of noise at a rate of 2\,MPixel/s, for a fraction of the power consumption\cite{porelMCRCspie2022, porelMCRCspie2024}.

\subsection{MCRC ASIC DEBUGGING FEATURES}
\label{sec:MCRC_debugging}

\begin{figure} [ht!]
   \begin{center}
   \begin{tabular}{c}
   \includegraphics[height=7cm]{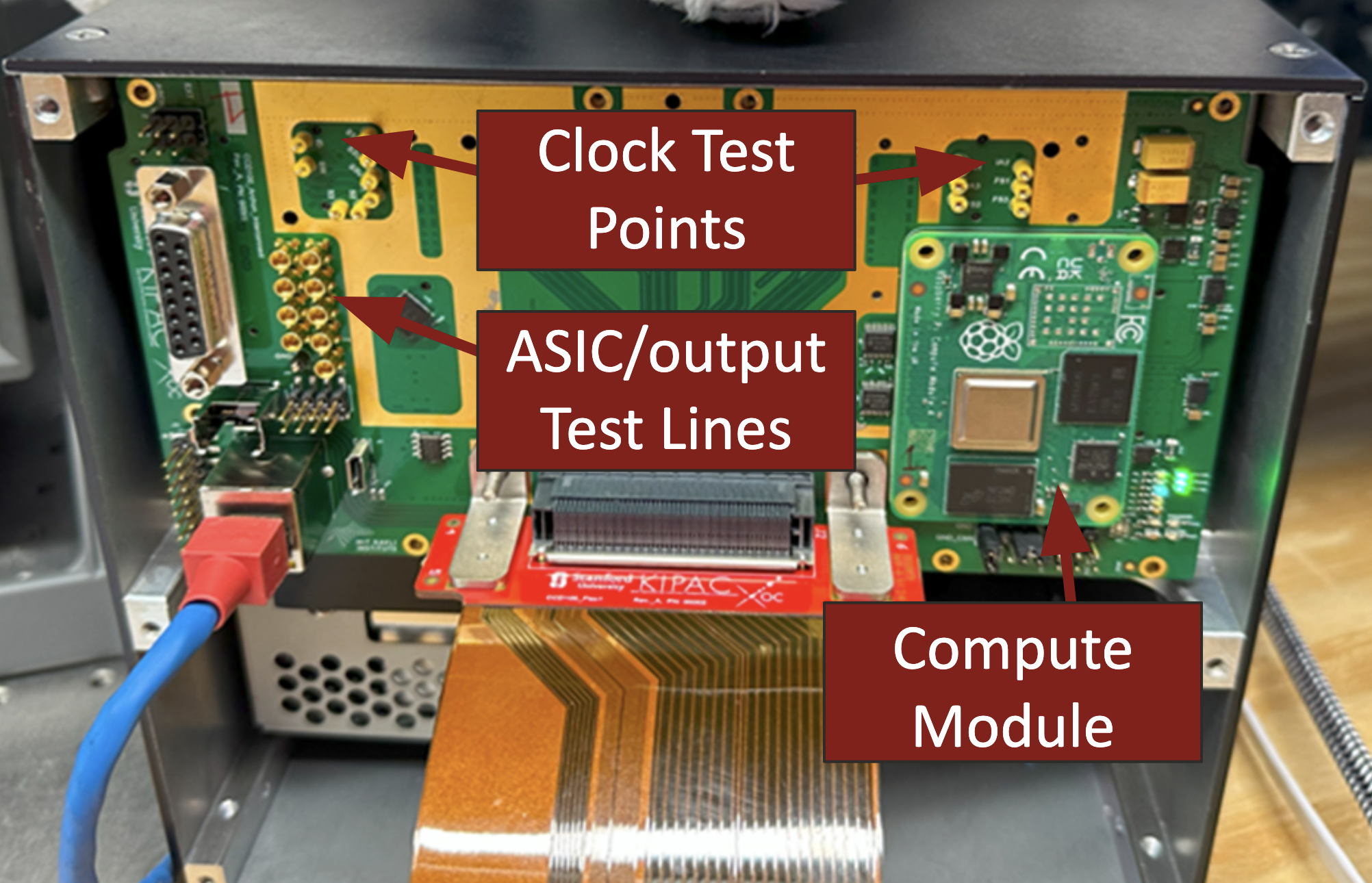}
   \end{tabular}
   \end{center}
   \caption[] 
   { \label{fig:interconnect} 
   Archon interconnect board for monitoring of CCD clock pulses with dedicated test points and for MCRC test pulse injection and output measurements with coax connectors. A Raspberry Pi compute module enables programming and monitoring of the MCRC ASIC.
}
\end{figure}

\noindent For the CCID-100 test system specifically, we designed the Archon interconnect board to augment the Archon controller with various diagnostic tools, including analog monitoring of CCD clocks, biases, currents, and output signals, as well as access to the dedicated MCRC V1 testing network internal to the ASIC. This offers the ability to test input connectivity and the source-follower detector output bias voltage on each channel. Further, it enables direct injection of test pulses and measurements of the readout chain response, and it runs diagnostics in-situ. Figure~\ref{fig:interconnect} shows the Archon interconnect board mounted on the Archon controller with labels for the clock test points, MCRC V1 test network access points, and CCD output coaxial connectors. The figure also shows the embedded compute module based on the Raspberry Pi hardware which is used for independent operation from the Archon to avoid interference with its digital signal processing (DSP) core. In addition, the compute module enables the concurrent usage of these diagnostic tools and the supporting command-line executable software for direct bidirectional communication and programming of the MCRC-V1 ASIC as well as real time monitoring and logging of voltages, currents and temperatures. 

\section{NOISE CHARACTERIZATION \& BIAS OPTIMIZATION}
\label{sec:bias_scan}

\begin{figure} [ht!]
   \begin{center}
   \begin{tabular}{c}
   \includegraphics[width=0.95\textwidth]{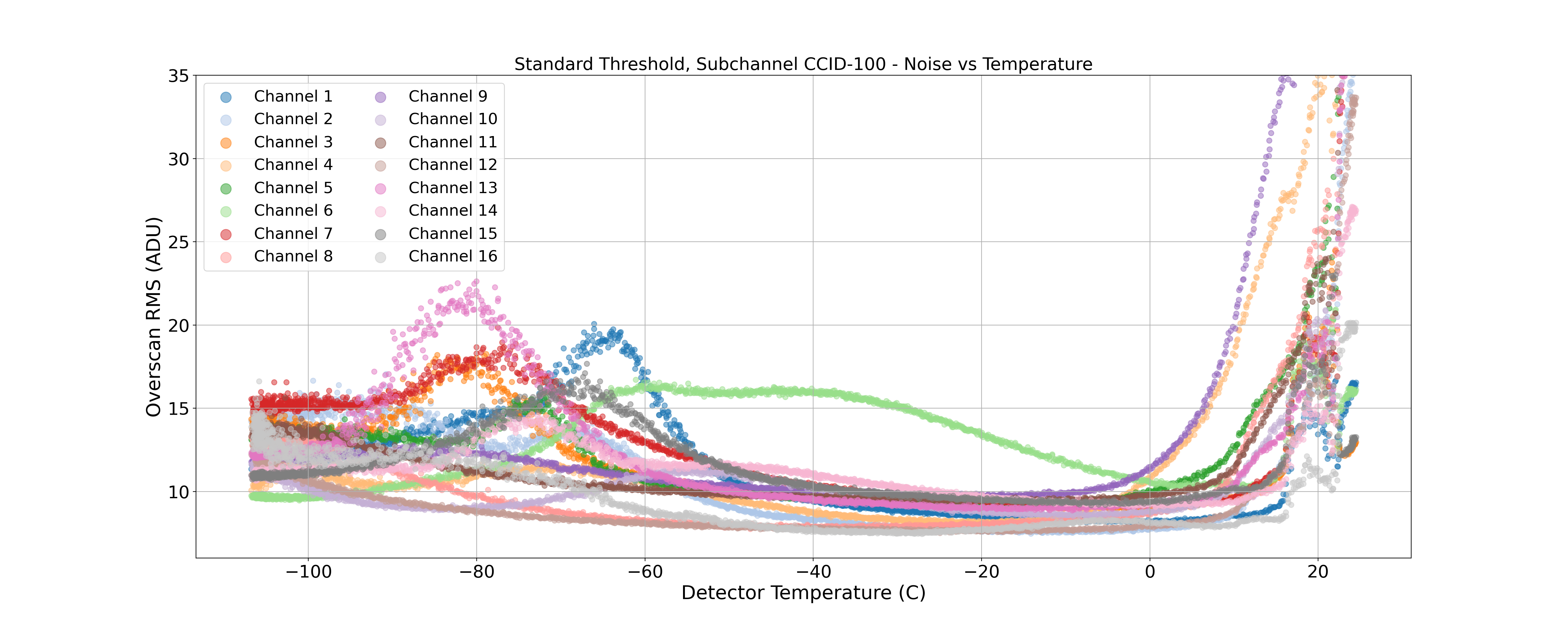}
   \end{tabular}
   \end{center}
   \caption[] 
   { \label{fig:noise_curveW6} 
   Read noise in ADU of each channel of the W6 CCID-100 as a function of temperature, taken as the detector was cooling. 
}
\end{figure}

\begin{figure} [ht!]
   \begin{center}
   \begin{tabular}{c}
   \includegraphics[width=0.95\textwidth]{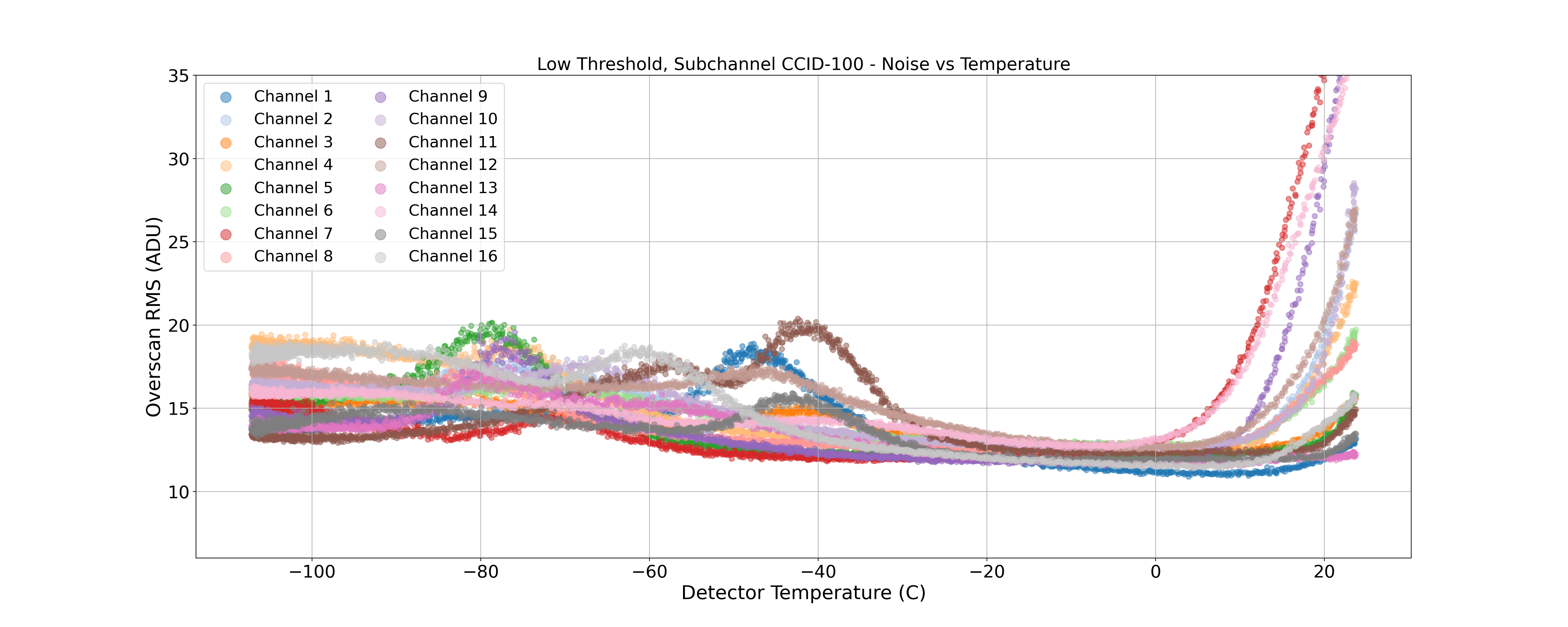}
   \end{tabular}
   \end{center}
   \caption[] 
   { \label{fig:noise_curveW16} 
   Read noise in ADU of each channel of the W16 CCID-100 as a function of temperature, taken as the detector was cooling. 
}
\end{figure}

\noindent The W6 and W16 detectors were cooled in the beamline to temperatures of approximately $-105\,\rm ^{\circ}C$. Image data were taken to monitor the read noise of each of the 16 output channels during cooling. The resulting noise curves for the W6 and W16 detectors are presented in Figures~\ref{fig:noise_curveW6} and \ref{fig:noise_curveW16}, respectively. The read noise is given in analog-to-digital units (ADUs). In this case, an ADU is equal to $4/2^{16}$ V or 61 $\mu$V. It can be seen that in both devices, as the temperature drops, the read noise reaches a minimum at $\sim -5^{\circ}\rm C$ to $-20^{\circ}\rm C$, but then increases and peaks in many channels at lower temperatures. This phenomenon has also been observed in the smaller-scale, MIT-LL CCID-93 detectors, which use the same output stage architecture and process technology\cite{Stueber2026InPrep}. Although the exact cause of this is not yet known, it has been seen that the output stage transistors of CCDs can produce such resonant noise peaks from 1/f noise induced by charge carrier trapping and detrapping\cite{Kandiah10.1109, Kandiah1991600, Janesick01}. Modeling of the temperature dependence of trap-induced 1/f noise produces resonant peaks consistent with those seen in the CCID-93 and CCID-100 devices. Due to this resonance behavior at low temperatures, which was especially prevalent in the W6 detector, we found that best detector performance was achieved for that device at $-40\,\rm ^{\circ}C$. Thus, we proceeded to optimize the performance of the W6 detector at this temperature, while optimizing the W16 detector at $-100\,\rm^{\circ}C$.

\begin{figure} [ht!]
   \begin{center}
   \begin{tabular}{c}
   \includegraphics[height=12cm]{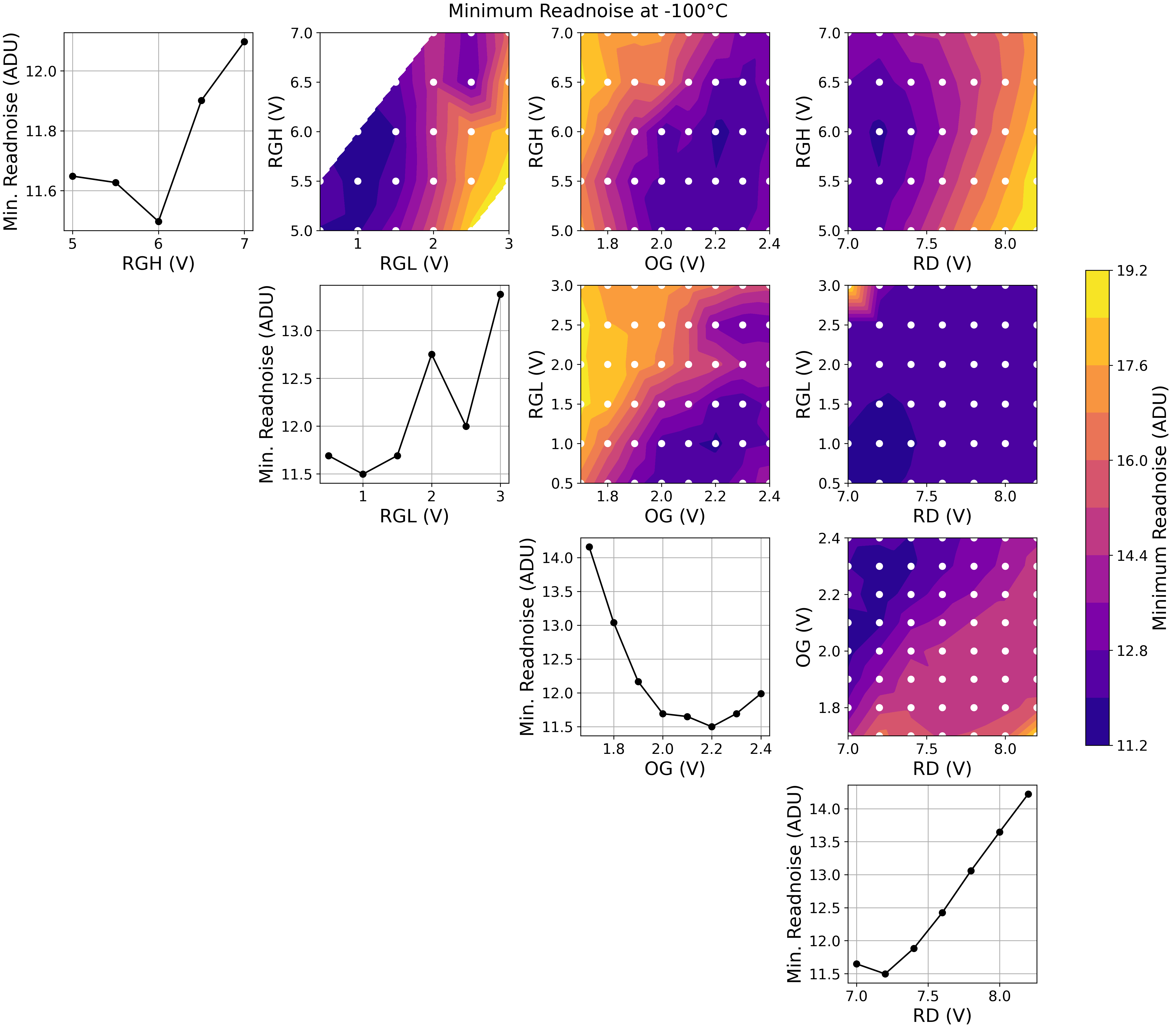}
   \end{tabular}
   \end{center}
   \caption[] 
   { \label{fig:bias_scan} 
   Triangle plots summarizing the minimum read noise achieved for all bias parameter RD, OG, RGH, and RGL combinations. Diagonal plots trace the minimum read noise of each individual parameter, obtained by minimizing over all combinations of the other three parameters. Off-diagonal contour plots linearly interpolate the minimum read noise achieved at different combinations of two parameters after minimizing over the other two.
}
\end{figure}

The resonant noise feature in the CCID-93 devices is sensitive to the biases of the output stage reset gate (RG), reset drain (RD), and output gate (OG)\cite{Stueber2026InPrep}, which are labeled on the schematic in Figure~\ref{fig:output}. Following the procedure outlined in [\citenum{Stueber10.1117}], we perform a bias point optimization by scanning RD, OG, and the high- and low-state voltages of RG, RGH and RGL, and record the read noise in ADU averaged across all 16 channels. Figure~\ref{fig:bias_scan} presents triangle graphs of the optimal read noise for the W16 CCID-100 detector at $-100\,\rm^{\circ}C$ in the range of values studied for RGH, RGL, OG and RD. In each diagonal plot, the recorded noise values are the minima at the given bias parameter when minimizing over all combinations of the other three parameters. The off-diagonal plots are two-dimensional, linearly interpolated contour plots that trace the minimum read noise for each combination of two given parameters, obtained by minimizing over the remaining two. White points indicate the bias values that were part of the scan, while gaps in the contours indicate combinations of parameter values that were not allowed by the scan and for which no data were collected. 

The bias optimization is able to complete a scan of $\sim\!1300$ parameter combinations in just under 3 hours. Table~\ref{tab:optimized_biases} summarizes the optimal bias points of the W6 and W16 detectors found by scans at a 2\,MPixel/s readout speed, at their respective temperatures of $-40\,\rm^{\circ}C$ and $-100\,\rm^{\circ}C$. The optimal biases for each detector are notably distinct; the noise and gain behavior of these devices may vary broadly across temperatures, devices, and even on an output-by-output basis. The bias scanning algorithm streamlines the process of optimizing detector performance under any set of conditions, which is otherwise complicated by the often complex and non-uniform nature of multi-channel devices.

\begin{table}[ht!]
    \caption{Optimized RGH, RGL, RD, and OG bias points in the W6 and W16 CCID-100 devices, for a readout speed of 2\,MPixel/s.}
    \label{tab:optimized_biases}
    \begin{center}
    \begin{tabular}{|c|c|c|c|c|c|}
    \hline
    \textbf{Device} & \textbf{Temp ($^{\circ}\rm C$)} & \textbf{RGH (V)} & {\textbf{RGL (V)}} & {\textbf{RD (V)}} & {\textbf{OG (V)}} \\
    \hline
    W6 & -40 & 6.5 & 1.5 & 8.1 & 1.9 \\
    W16 & -100 & 6.0 & 1.0 & 7.2 & 2.2 \\
    \hline
    \end{tabular}
    \end{center}
\end{table} 

\begin{figure} [ht!]
   \begin{center}
   \begin{tabular}{c}
   \includegraphics[height=6cm]{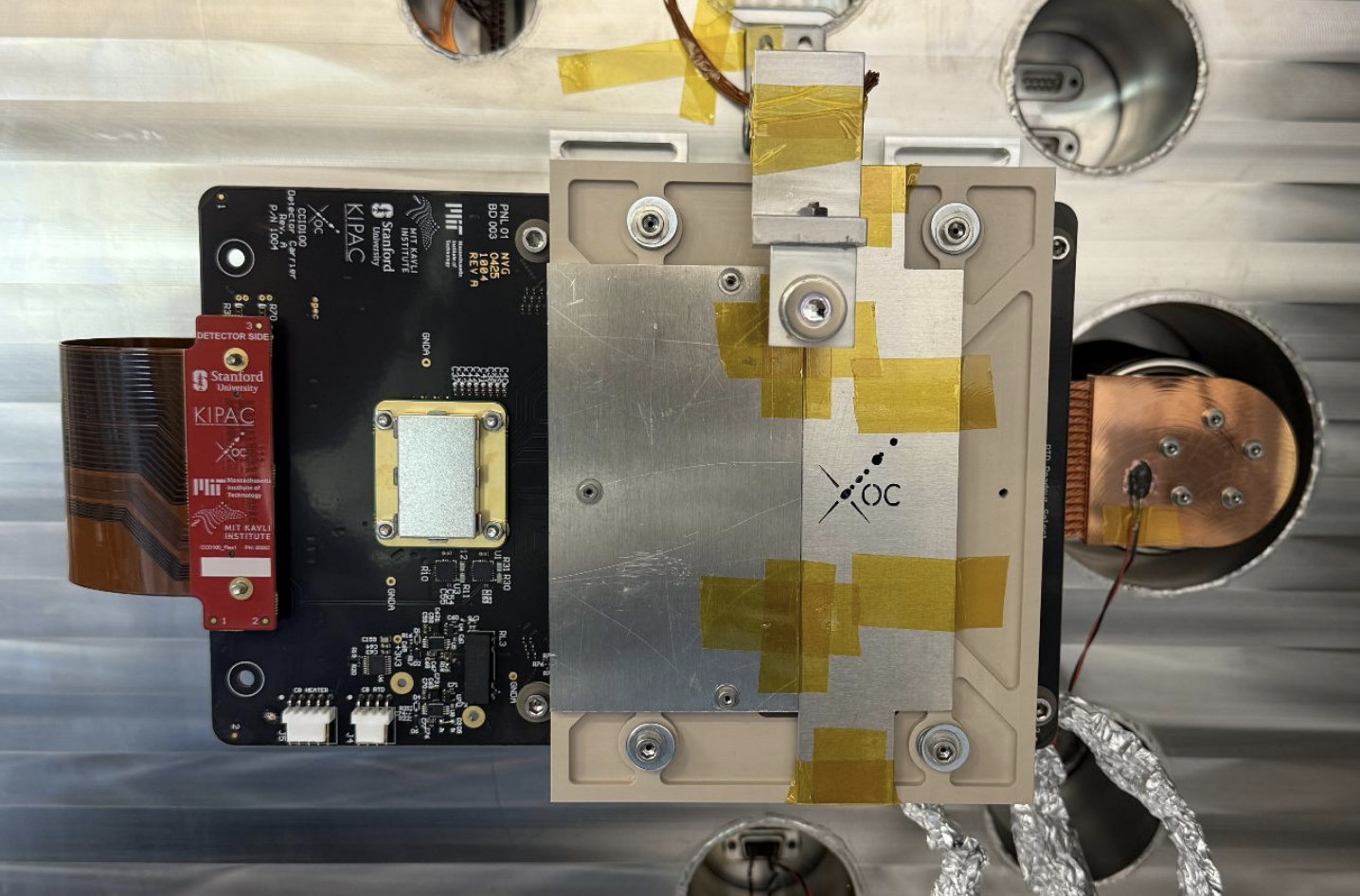}
   \includegraphics[height=6cm]{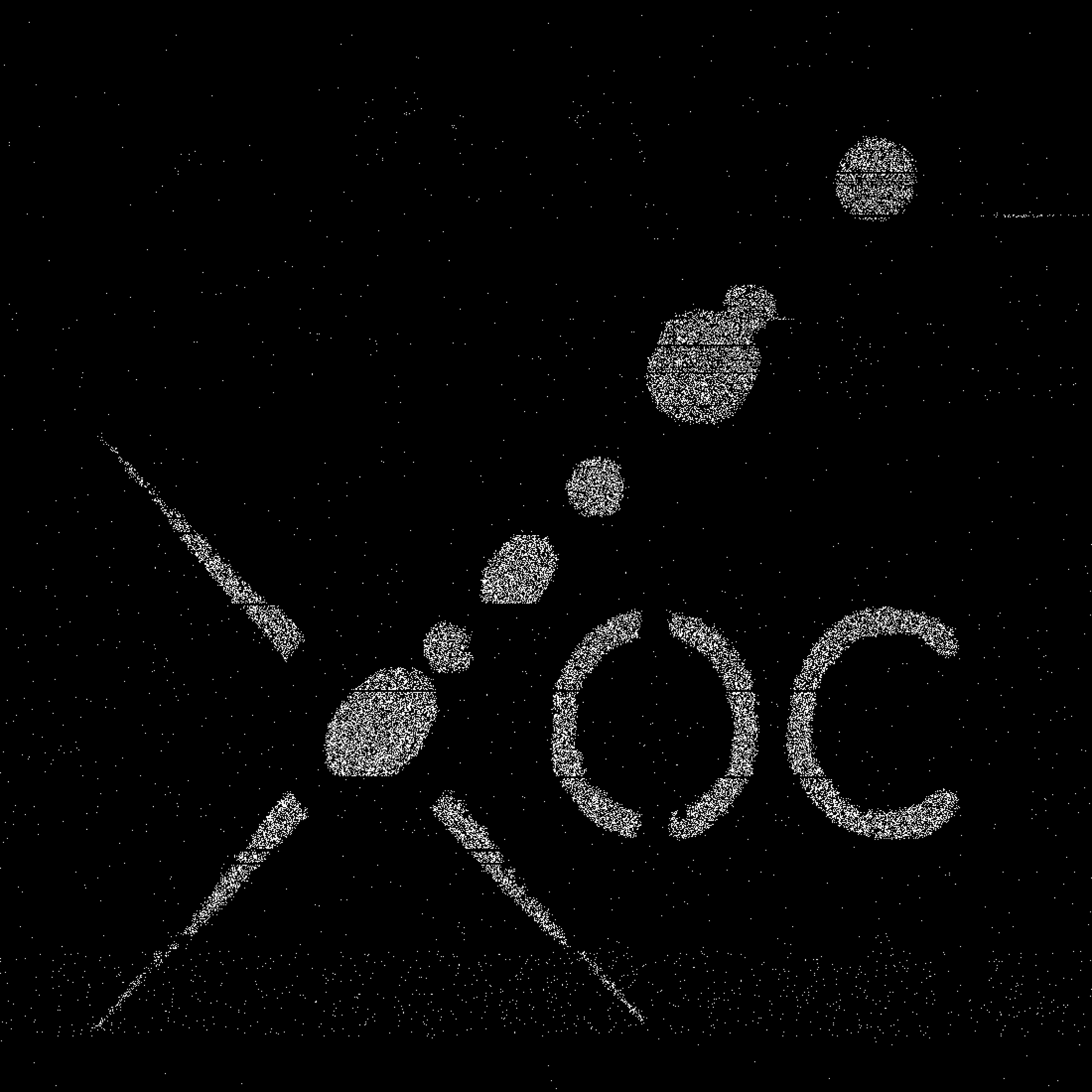}
   \end{tabular}
   \end{center}
   \caption[] 
   { \label{fig:g0_reconstructed} 
   {\it Left:} Detector mounted in the Gen 1.0 XOC X-ray beamline with ``XOC'' cutout aluminum cover secured over top of the imaging area.
   {\it Right:} Reconstructed CCID-100 image of single pixel events produced by Titanium (4.5\,keV) fluorescane photons.
}
\end{figure}

With the detector bias optimized, we inserted an aluminum mask with an ``XOC'' logo cutout in front of the detector (Fig.~\ref{fig:g0_reconstructed}, left) to produce an event reconstructed image of the logo with the CCD operating in full frame readout mode. That is, the CCD was exposed for 500\,ms to 4.5\,keV Titanium fluorescence photons, after which the image data were transferred to the frame store and then read out while another image began integrating. We then identified the single pixel events in each image, and reconstructed pixel-by-pixel the locations of each event. The resulting reconstructed image is seen on the right of Figure~\ref{fig:g0_reconstructed}. 

\section{RESULTS}
\label{sec:results}

\begin{figure} [ht!]
   \begin{center}
   \begin{tabular}{c}
   \includegraphics[height=6.5cm]{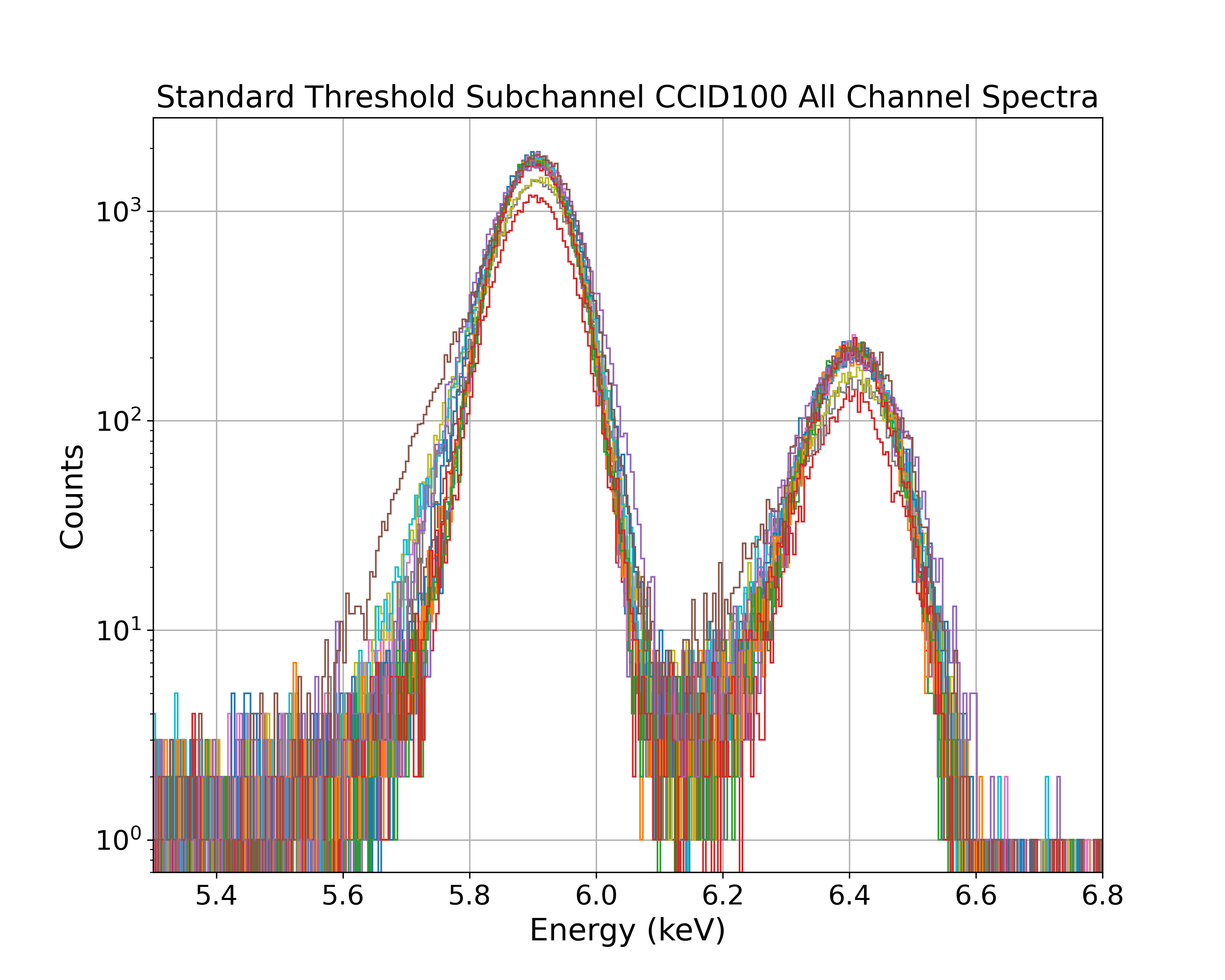}
   \includegraphics[height=6.5cm]{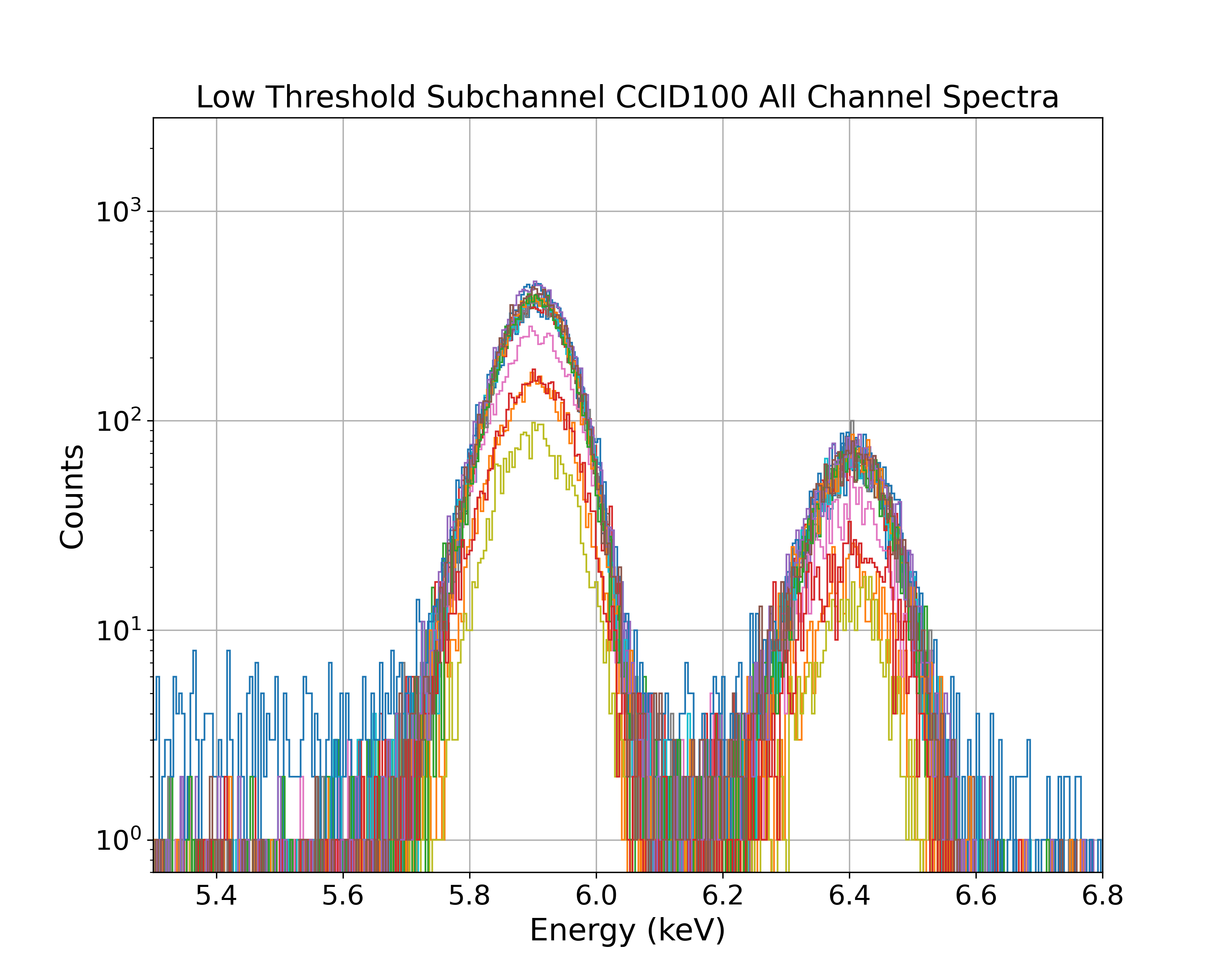}
   \end{tabular}
   \end{center}
   \caption[] 
   { \label{fig:spectra} 
   {\it Left:} All 16 channel spectra obtained from the standard-threshold, subchannel (W6) CCID-100 detector, operated at 2\,MPixel/s serial readout speed at $-40\,\rm^{\circ}C$. {\it Right:} All 16 channel spectra obtained from the low-threshold, subchannel (W16) CCID-100 detector, operated at 2\,MPixel/s serial readout speed at $-100\,\rm^{\circ}C$. 
}
\end{figure}

We present in Figure~\ref{fig:spectra} the Fe-55, single pixel event spectra in each channel of the W6 CCID-100 at $-40\,\rm^{\circ}C$ (left), and the W16 CCID-100 at $-100\,\rm^{\circ}C$ (right) at their respective optimal biases given in Table~\ref{tab:optimized_biases}. Both spectra were obtained at a serial readout speed of 2\,MPixel/s and parallel transfer speed of approximately 130\,kHz. The frame rate achieved was $\sim\!6.7$ frames/s. 

All channels across both detectors are cosmetically clean and operational, producing reasonable spectra at high frame rates while maintaining low noise levels. Many of the W6 and W16 channels would meet the baseline requirements for a mission like AXIS, which specified 3\,electron read noise and less than 150\,eV FWHM on the Fe-55 5.9\,keV line.

In the W6 CCID-100 there are:
\begin{itemize}
    \item 11 Channels with noise $<3.0$\,electrons.
    \item 15 Channels with noise $<3.2$\,electrons.
    \item 12 Channels with single pixel event spectra with $<150$\,eV FWHM at 5.9\,keV.
\end{itemize}

In the W16 CCID-100 there are:
\begin{itemize}
    \item 4 Channels with noise $<3.0$\,electrons.
    \item 10 Channels with noise $<3.5$\,electrons.
    \item 14 Channels with single pixel event spectra with $<150$\,eV FWHM at 5.9\,keV.
\end{itemize}

\section{CONCLUSIONS}
\label{sec:conclusions}

The CCID-100 devices are the first multi-megapixel, 16-channel imaging CCDs to demonstrate speed and noise performance at the levels required of next generation, wide-field, high-resolution X-ray imaging observatories. All channels of two front-illuminated test devices, the W6 standard threshold, subchannel CCID-100 and the W16 low threshold, subchannel CCID-100, were fully operational and produced spectra. Many of the channels on these first devices met the noise and energy resolution requirements outlined by the AXIS probe mission concept. Rapid, parallelized readout of all 16 channels was enabled by dual MCRC V1 ASIC devices, and was further enhanced with the implementation of thorough diagnostics and testing networks. Testing and characterization of these large-scale, multi-channel devices has been streamlined and simplified via the implementation of automated bias scanning algorithms to efficiently optimize new devices. Though originally developed for the AXIS concept, the demonstrated speed, noise, and spectral performance of the CCID-100 make it a suitable, mission-ready imager for other, future strategic X-ray missions.  

\appendix

\acknowledgments 

We acknowledge support from the NASA Astrophysics Probe Explorer (APEX) program under contract number 80GSFC25CA019, the NASA Astrophysics Research and Analysis (APRA) program under grant number 80NSSC22K1921, and the NASA Strategic Astrophysics Technology (SAT) program under grant number 80NSSC23K0211.


\bibliographystyle{spiebib} 

\end{document}